RESEARCH-ARTICLE

# IoT-enabled Stability Chamber for the Pharmaceutical Industry

**NITOL SAHA**, University of South Carolina, Columbia, SC, United States

**MD MASRUK AULIA**, Military Institute of Science and Technology, Dhaka, Bangladesh

**DIBAKAR DAS**, Bangladesh University of Engineering and Technology, Dhaka, Dhaka, Bangladesh

**MD MOSTAFIZUR RAHMAN**, Khulna University of Engineering and Technology, Khulna, Bangladesh

# IoT-enabled Stability Chamber for the Pharmaceutical Industry

Nitol Saha*
University of South Carolina
Columbia, South Carolina, USA
nsaha@email.sc.edu

Md Masruk Aulia
Military Institute of Science & Technology (MIST)
Dhaka, Bangladesh
masruk.aulia@gmail.com

Dibakar Das
Bangladesh University of Engineering & Technology (BUET)
Dhaka, Bangladesh
dibakarsunny@yahoo.com

Md. Mostafizur Rahman
Khulna University of Engineering & Technology (KUET)
Khulna, Bangladesh
mostafiz.kuetme@gmail.com

## Abstract

A stability chamber is essential for pharmaceutical facilities to test the stability and quality of products over time by exposing them to different environmental conditions. This paper introduces an IoT-enabled stability chamber designed for the pharmaceutical industry. We constructed four stability chambers by leveraging the existing infrastructure within a manufacturing facility. Each chamber is controlled using a state-of-the-art Proportional Integral Derivative (PID) system based on the Siemens S7-1200 PLC. The Siemens WinCC Runtime Advanced platform, compliant with FDA 21 CFR Part 11, was used for visualizing chamber data. Additionally, an Internet of Things (IoT) application was developed to remotely monitor sensor data through any client application. This research aims to enhance the performance of traditional stability chambers by integrating IoT functionalities, making them more cost-effective and user-friendly.

## 1 Introduction

The pharmaceutical industry is an ever-growing industry that is continuously striving to enhance product quality, efficacy, and safety. The storage and maintenance of drugs is one of the most critical aspects of it. The efficacy and safety of pharmaceutical products can be greatly influenced by the loss of potency during storage of the product [1]. All locations where medications and supplies are kept on the property need to maintain appropriate environmental management, which includes appropriate lighting, temperature, and humidity as well as hygienic conditions, ventilation, and segregation [2]. Stability chambers play a pivotal role in this process by providing controlled environments to simulate various temperature and humidity conditions, mimicking real-world scenarios encountered during storage and transportation. A study conducted in Saudi Arabia with 1,000 participants found that over 30% were unable to identify the most appropriate storage conditions, and 10% were unaware that storage conditions affect a product's shelf life [3]. Temperature and humidity are crucial for storing other products like seeds. Research has shown that soybean seeds can remain viable for two seasons if stored in conditions where the temperature is kept at 10℃ and the humidity is less than 40% [4]. This leads us to the fact that the stability studies of the product are important as they have a critical impact on the condition of the products.

Stability studies are standard operating procedures to guarantee the preservation of pharmaceutical product safety over the course of its shelf life. Stability testing shows how a medication product's quality changes over time in response to different environmental variables [5]. Stability studies are also very important for the development and commercialization of a new medicine [6]. To have all these stability studies and for the maintenance of pharmaceutical products, stability chambers come into practice.

## 2 Literature Review

Stability chambers are a type of laboratory equipment used to simulate stable temperature and humidity-controlled conditions for product testing and storage purposes. Traditional stability chambers are effective, but they lack modern features like real-time remote monitoring and control functions which lead to potential risks to the product. With the integration of IoT, these difficulties have been overcome without a major increase in the manufacturing cost of these stability chambers.

The integrated system outlined in [7] can reliably provide information about the quality of packaged products throughout their storage period. This is achieved by using various sensors to monitor food quality and safety, recording parameters such as pathogen levels, gas concentrations, temperature, humidity, and storage duration. With the incorporation of IoT, this technology offers far more detailed information compared to traditional food inspection methods, which are typically limited to assessing weight, volume, color, and appearance. The system described is based on a straightforward yet effective approach to integrated food monitoring, utilizing IoT to create an interconnected network of devices. This allows for the integration of sensors and actuators, enabling them to share data across platforms and provide a comprehensive overview of the situation. Specifically, the system uses sensors for gas, temperature, and humidity to collect critical data for assessing the quality of packaged products. This data is wirelessly transmitted to a computer system, offering an interface where users can track the quality of products over time. The climate chamber discussed in [8] is a device located at the Technobothnia Education and Research Center which is used to simulate different climate conditions. A simulated environment is utilized to assess the performance of electrical equipment under varying temperature and humidity conditions. This climate chamber contains a specialized computer, referred to as the control PC, along with control software. Together, they manage and execute the specified simulations. The paper [9] presents a system designed for the calibration of thermohygrometers, which requires the ability to generate specific temperatures and relative humidity levels. To achieve this, a climate chamber was constructed to both generate and control these parameters. Temperature regulation is achieved using a thermoresistor, while relative humidity reduction is accomplished with a silica gel trap that recirculates the air inside the chamber using a vacuum pump. To increase relative humidity, an ultrasonic humidifier, also known as a fogger, is employed. The chamber is primarily made of acrylic, allowing for transparency and easy observation of the thermohygrometers' readings. The control of process variables, such as temperature and relative humidity, is managed by an Arduino board using PWM (pulse width modulation), resulting in precise control with variations of 0.3℃ and 2% relative humidity. The environmental chamber described in [10] was designed to house standard resistors under controlled temperature and humidity conditions during air calibration. The chamber maintains a temperature of 23℃ with a tolerance of ±2℃ using a proportional, integral, and derivative (PID) controller for precise heating. Relative humidity is kept at 35% with a tolerance of ±5%, controlled by aqueous salt solutions. The above-mentioned systems have been built for different products. However, pharmaceutical products are prone to be found out of desired properties if not maintained within the optimum temperature and relative humidity. For this purpose, an IoT-based stability chamber has been introduced in this paper with the inclusion of sensors, actuators, valves, PID control, and 21 CFR part 11 compliant for electronic data integrity [11]. This system enables continuous monitoring and precise control of environmental parameters such as temperature, and humidity and can proactively identify deviations from desired conditions, allowing for timely intervention to prevent product degradation or loss. There are also features like enhanced monitoring, user-level authorization, audit trail, digital alarm notification, and recording of data. This stability chamber is focused on the pharmaceutical sector as the pharmaceutical drugs are designed to be kept in specific conditions.

*Corresponding Author

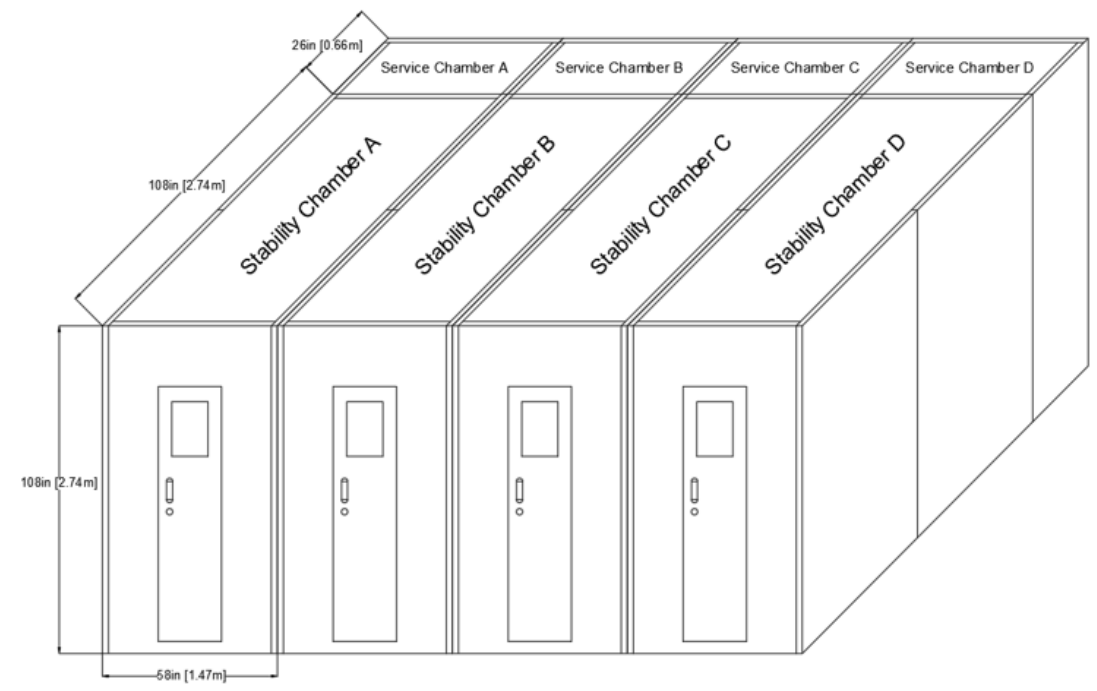

**Figure 1: Schematic Diagram of Stability Testing Chamber**

## 3 MECHANICAL SYSTEM DEVELOPMENT

The developed stability testing chamber consists of four different chambers denoted by A, B, C, and D, having different environmental conditions in each cell which is illustrated in Figure 1. in the Cartesian coordinate system. The width of each chamber is 1.47 m with a length and height of 3.4 m and 2.74 m respectively. The total volume taken by the stability testing chamber is 13.76 $m^3$.

Temperature (T) and Relative Humidity (RH) are the primary parameters of environmental conditions, which are maintained very precisely in each chamber. The environmental conditions maintained in the stability testing chamber are listed in Table 1.

These four chambers are equal in size, and each having an internal volume of 9.57 $m^3$ for testing purposes. Access to these chambers is facilitated through front doors, with each chamber containing seven racks of different sizes. These racks offer the flexibility to accommodate up to 78 trays for the storage of manufactured samples intended for long-term testing. Moreover, users have the convenience of easily repositioning trays within the racks for accessibility. Furthermore, a dedicated passage is provided to allow comfortable movement for personnel tasked with product handling within the chambers. The sandwich panel of the stability testing chamber is constructed using SS304 and expanded polystyrene (with a density of 20 $kg/m^3$). Air, serving as the working fluid within these chambers, is circulated by a twin blower. This blower intakes air from the rear part of the chamber, while the airflow returns through a duct positioned beneath the racks. The duct features seven openings beneath the racks for air recirculation to the blower. The system operates at a static pressure of 125 Pa. According to the performance graph of the blower, the airflow is 2375 $m^3$/hour. The position of the door, rack and tray, duct, passage for movement, air flow direction, and the service chamber are depicted in Figure 2.

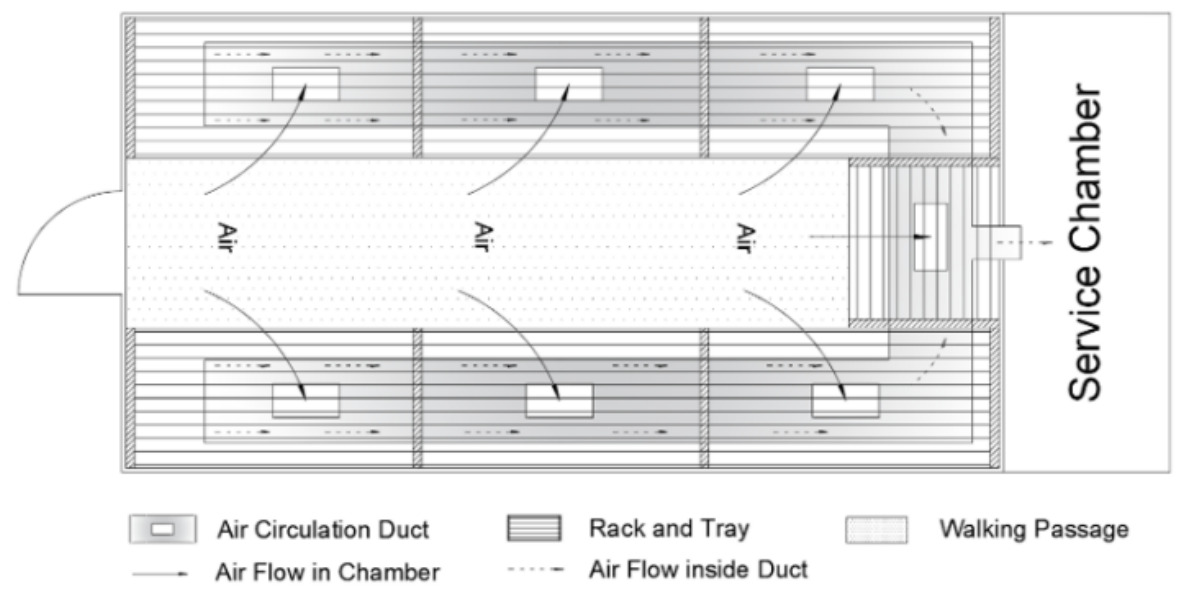

**Figure 2: Schematic Diagram of the Top view of the Stability Chamber**

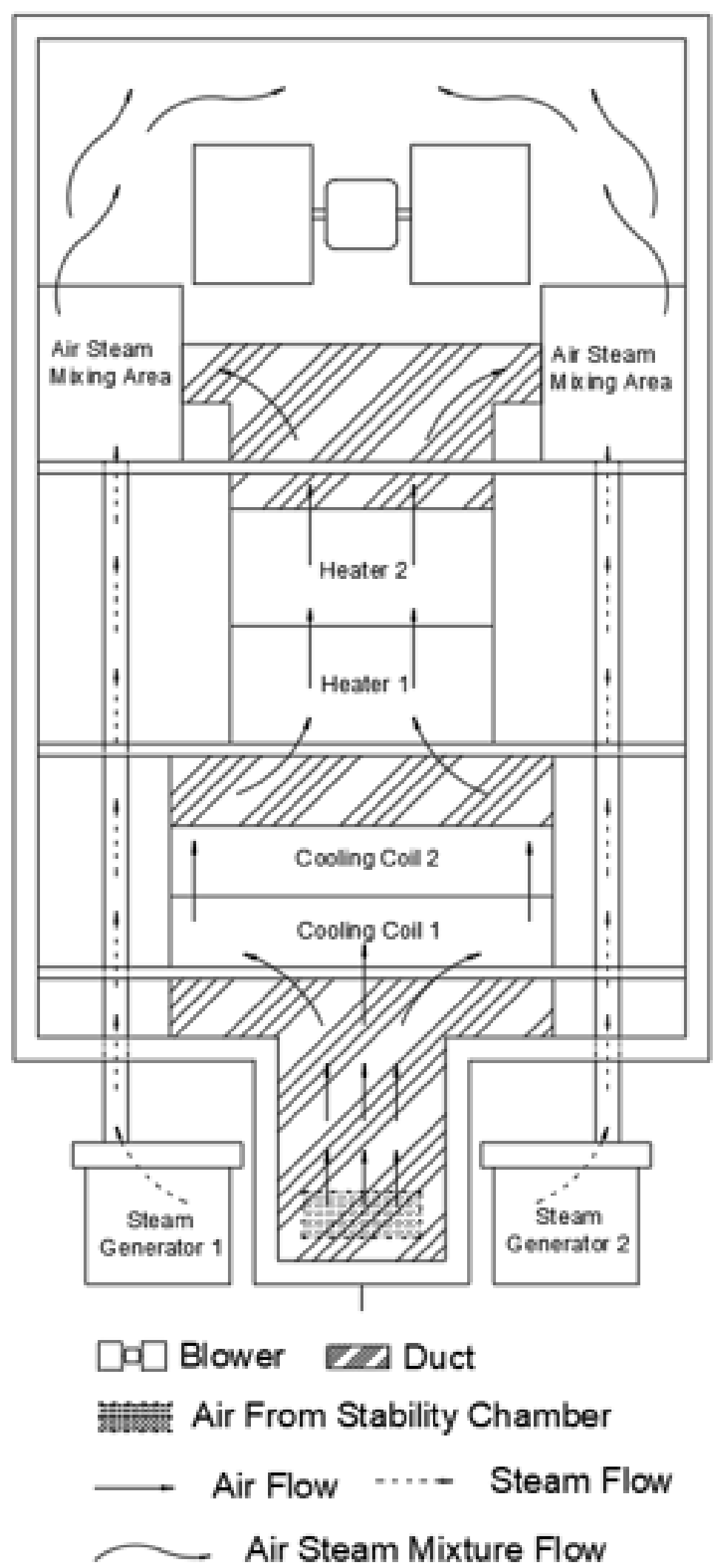

**Figure 3: Schematic Diagram of the Utility of the Stability Testing Chamber**

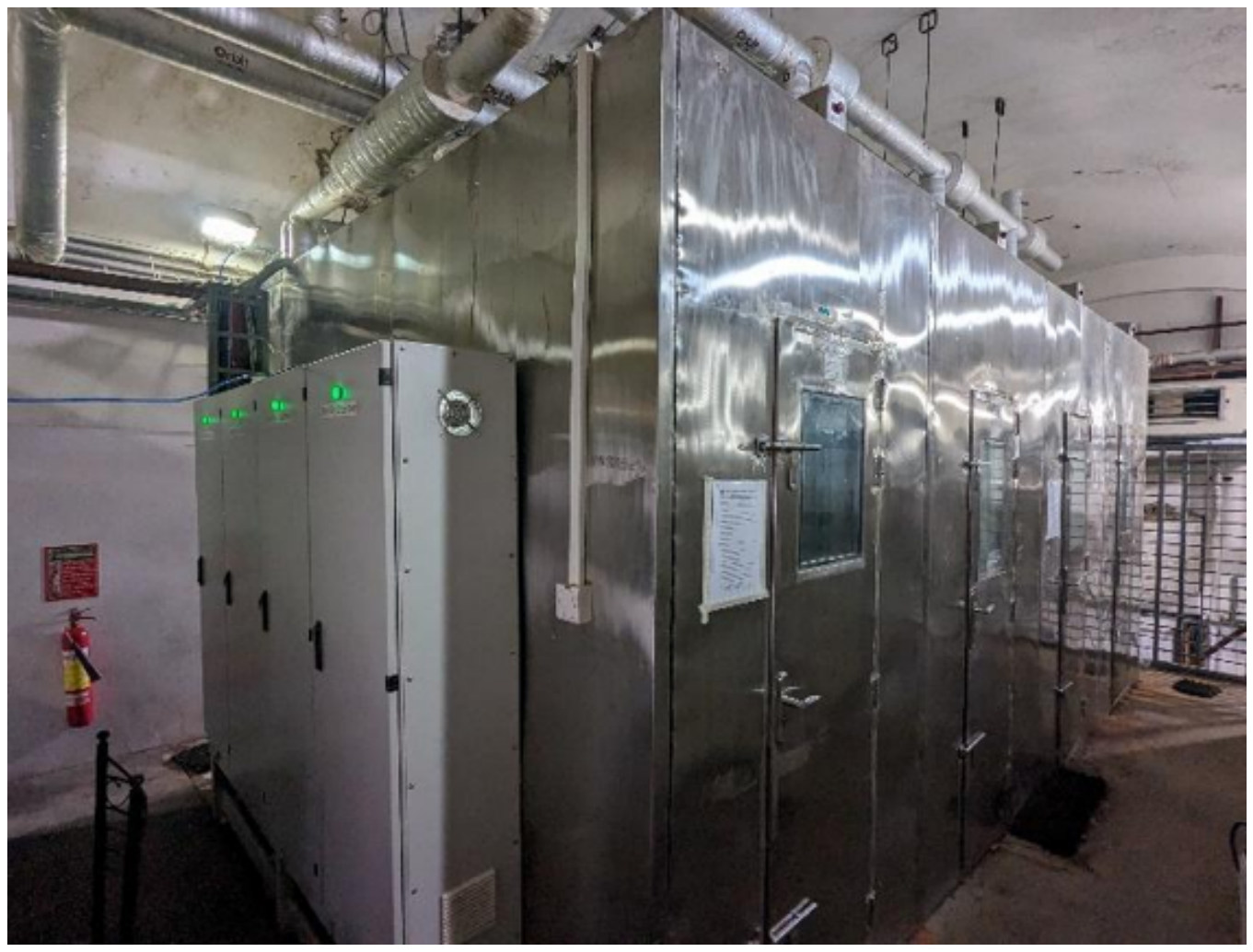

Figure 4: Schematic Diagram of the Utility of the Stability Testing Chamber

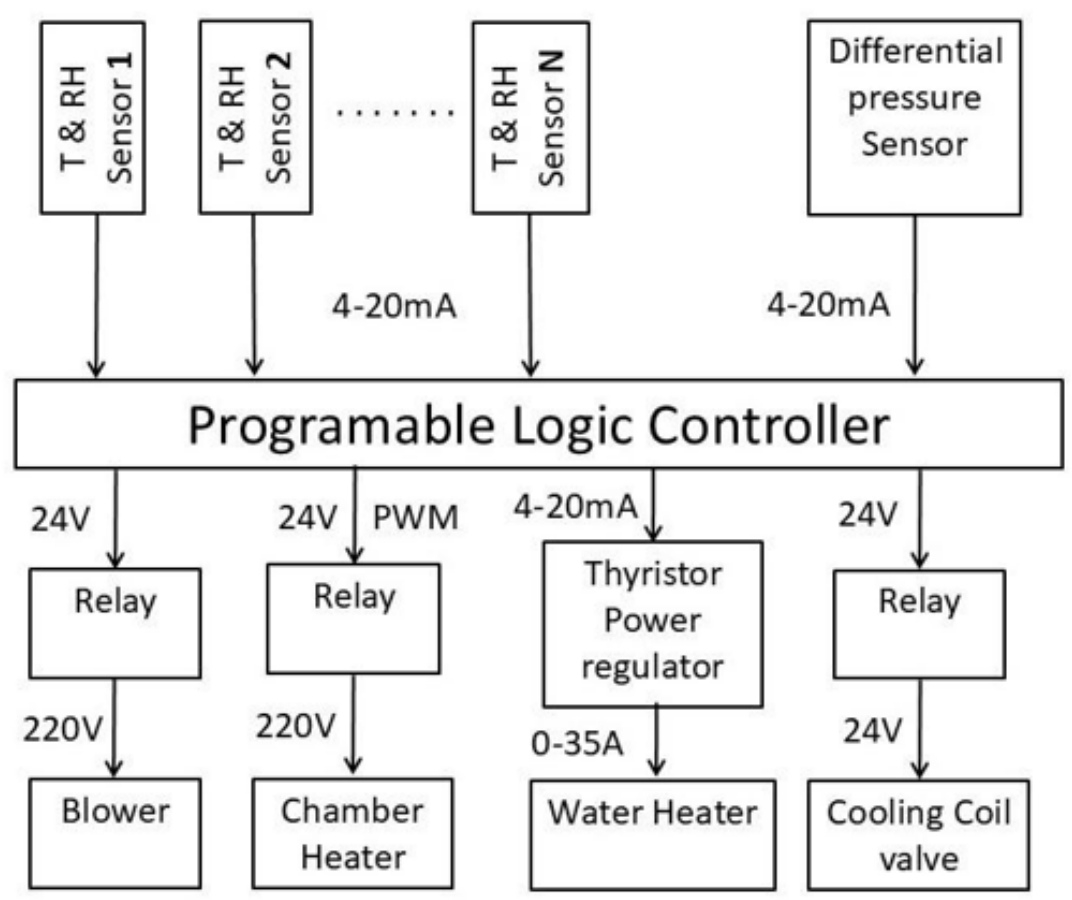

Figure 5: Schematic Diagram of the Control System

Table 1: Environmental Conditions of the Stability Testing Chamber in Each Compartment

| Chamber | Temperature (℃) | Relative Humidity (%) |
|---|---|---|
| A | 25 | 60 |
| B | 30 | 65 |
| C | 30 | 75 |
| D | 40 | 75 |

The return air undergoes cooling and heating processes within the service chamber before reentering the blower (Figure 3). The whole unit operates on a feedback mechanism wherein temperature fluctuations trigger the activation of either a cooling coil or a heater to maintain the desired temperature. The cooling coil is connected with a chilled water line, with coolant temperatures typically ranging between 7℃ and 8℃. A solenoid valve regulates the flow of

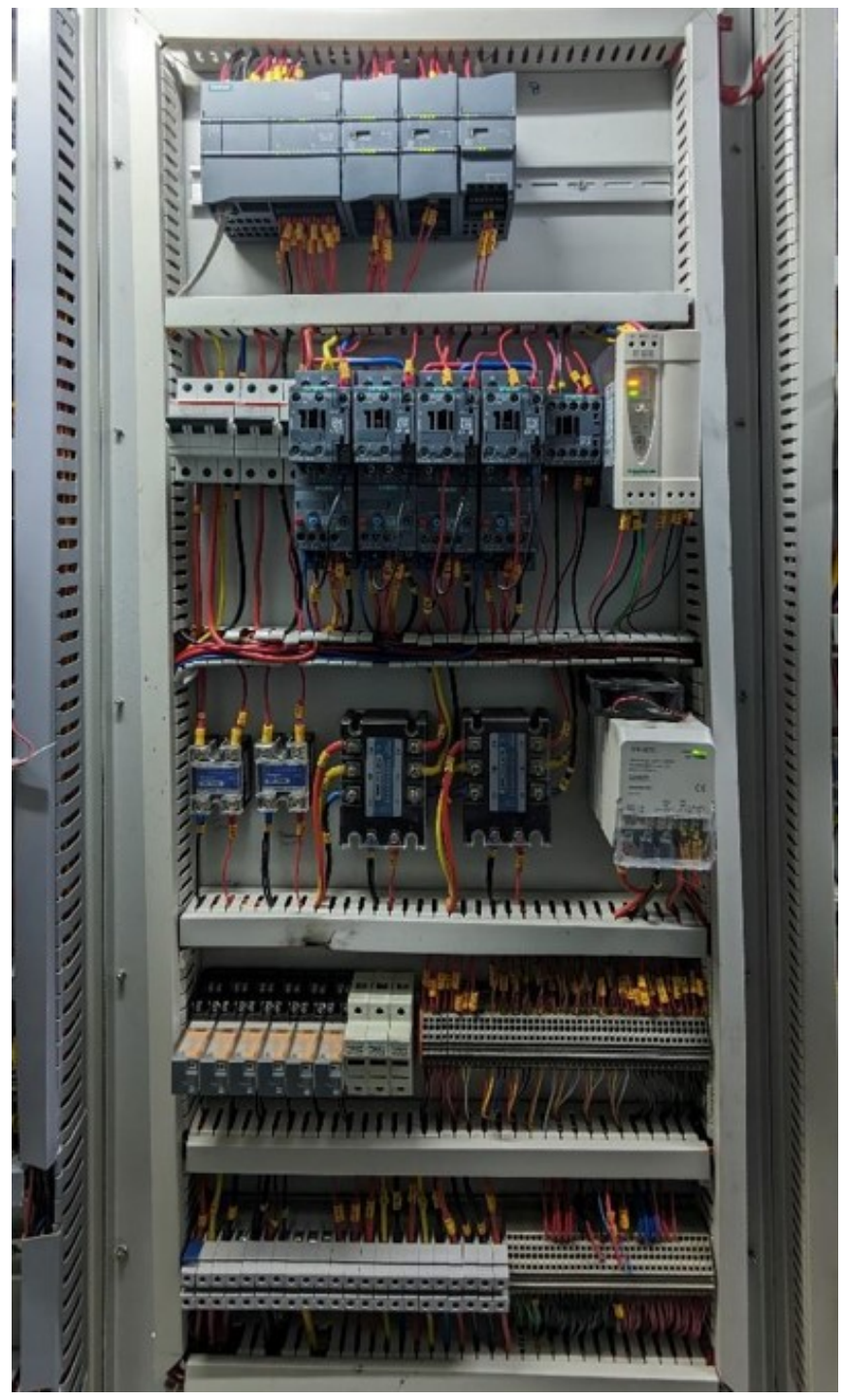

Figure 6: Control System of the Proposed System

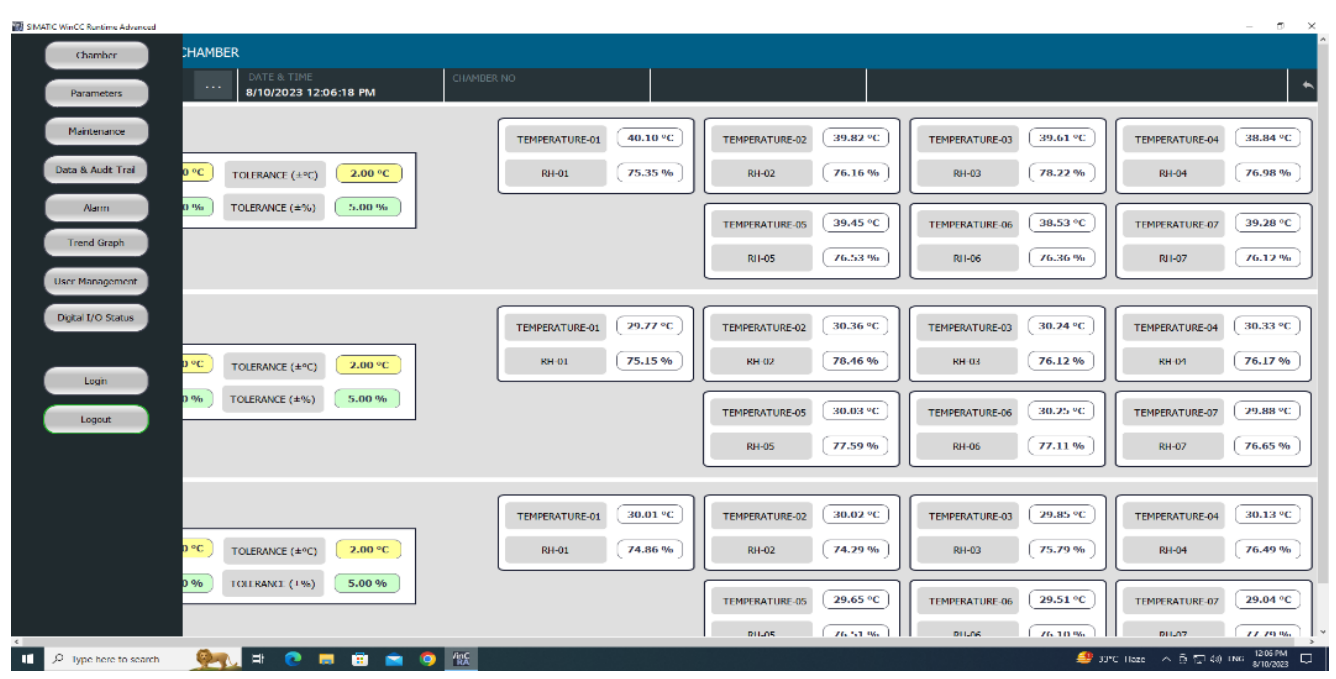

Figure 7: Chamber Monitoring Screen

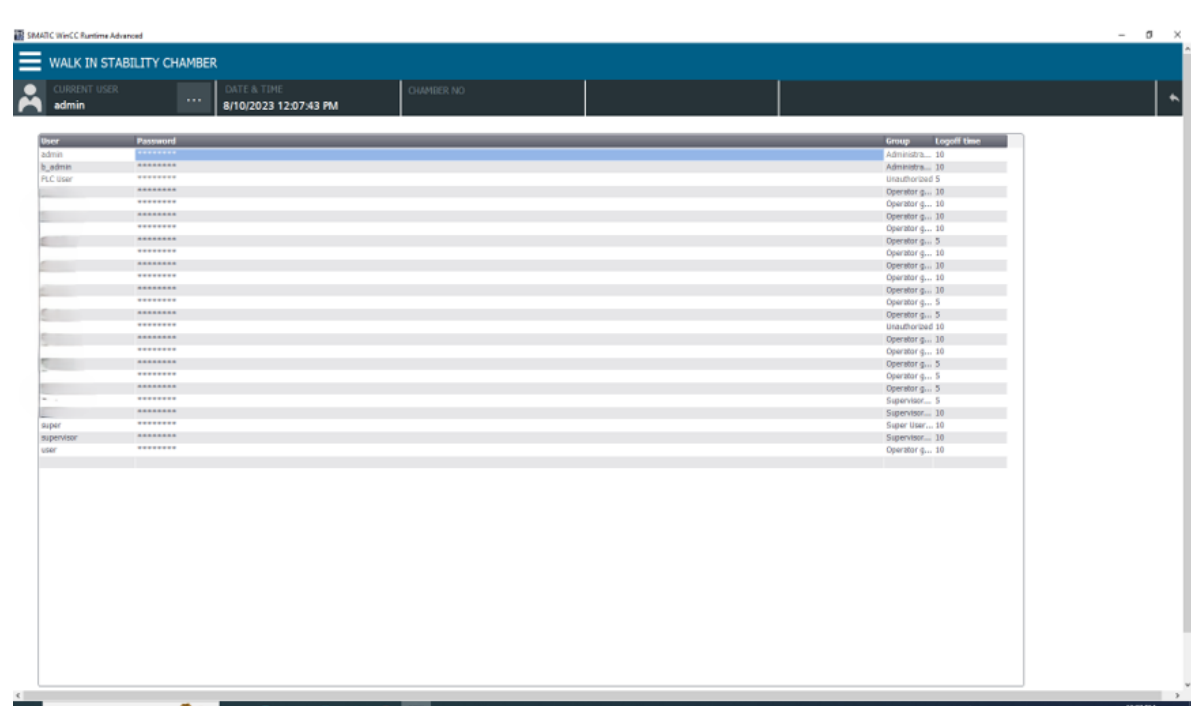

Figure 8: User Management Screen

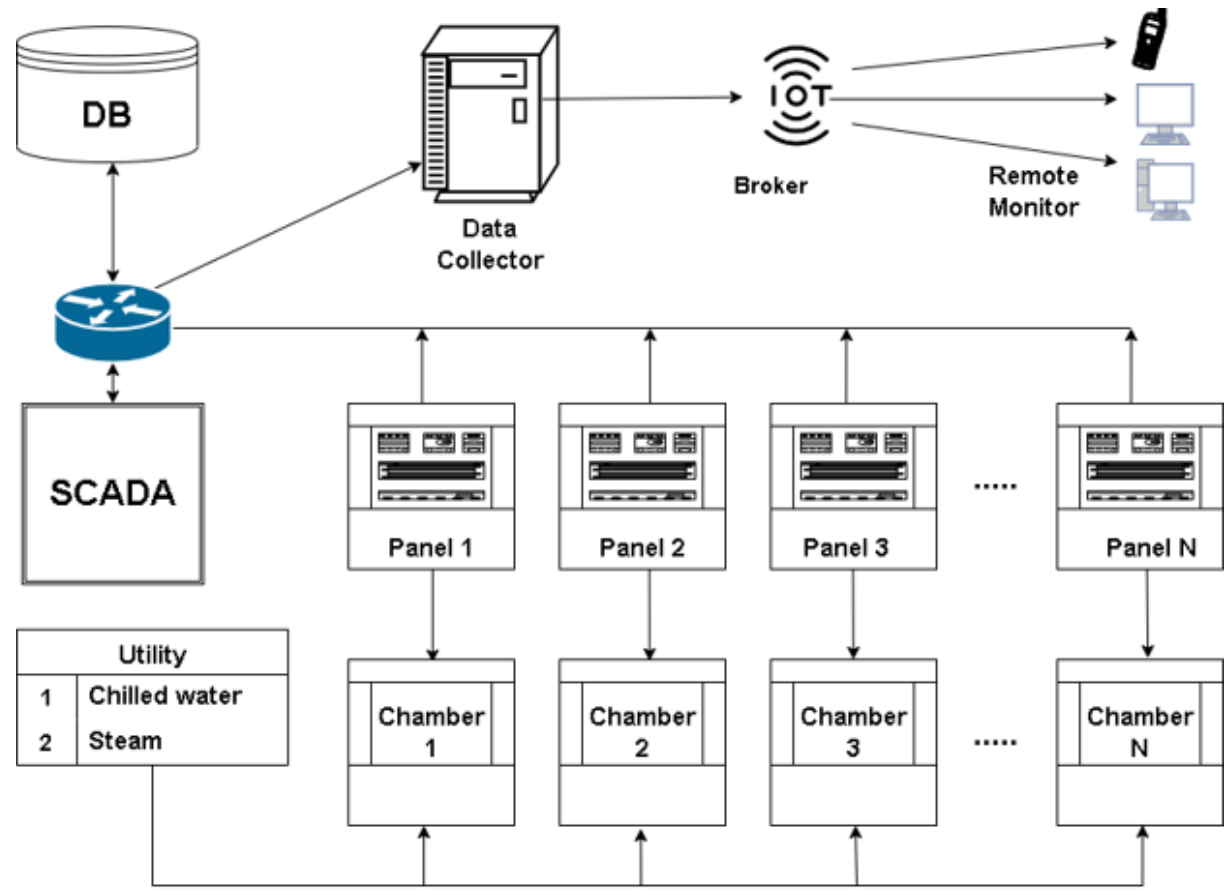

**Figure 9: IoT Architecture of the Proposed System**

Temperature & RH data of Walk-in Chamber-01

Data Report from 5/2/2024 3:30:00 PM to 5/3/2024 3:30:00 PM (Time Interval: 60 min)

| Serial | Formatted Time | T1 | RH1 | T2 | RH2 | T3 | RH3 | T4 | RH4 | T5 | RH5 | T6 | RH6 | T7 | RH7 |
|---|---|---|---|---|---|---|---|---|---|---|---|---|---|---|---|
| 1 | 5/2/2024 3:45:30 PM | 39.74 | 76.14 | 39.76 | 74.38 | 40.33 | 75.44 | 40.13 | 74.66 | 39.37 | 76.01 | 40.45 | 75.52 | 39.08 | 77.00 |
| 2 | 5/2/2024 4:45:30 PM | 39.79 | 76.12 | 39.64 | 74.51 | 40.41 | 75.59 | 40.15 | 74.69 | 39.36 | 76.03 | 40.51 | 75.64 | 39.08 | 77.15 |
| 3 | 5/2/2024 5:45:30 PM | 39.83 | 76.07 | 39.84 | 74.42 | 40.44 | 75.16 | 40.26 | 74.64 | 39.47 | 75.88 | 40.54 | 75.38 | 39.20 | 76.75 |
| 4 | 5/2/2024 6:45:30 PM | 39.86 | 75.89 | 39.87 | 74.06 | 40.45 | 75.38 | 40.30 | 74.40 | 39.48 | 75.88 | 40.56 | 75.21 | 39.19 | 76.61 |
| 5 | 5/2/2024 7:45:30 PM | 39.80 | 76.79 | 39.74 | 75.03 | 40.43 | 76.28 | 40.24 | 75.44 | 39.51 | 76.68 | 40.53 | 76.34 | 39.19 | 77.42 |
| 6 | 5/2/2024 8:45:30 PM | 39.73 | 76.14 | 39.86 | 74.50 | 40.43 | 75.64 | 40.27 | 74.76 | 39.47 | 76.11 | 40.56 | 75.65 | 39.19 | 77.16 |
| 7 | 5/2/2024 9:45:30 PM | 39.73 | 76.52 | 39.77 | 74.69 | 40.34 | 76.05 | 40.12 | 75.06 | 39.37 | 76.28 | 40.44 | 75.91 | 38.99 | 77.30 |
| 8 | 5/2/2024 10:45:30 PM | 39.65 | 76.16 | 39.76 | 74.52 | 40.27 | 75.91 | 40.12 | 75.03 | 39.41 | 76.30 | 40.41 | 75.84 | 39.05 | 77.00 |
| 9 | 5/2/2024 11:45:30 PM | 39.82 | 76.41 | 39.85 | 74.64 | 40.34 | 75.85 | 40.17 | 75.05 | 39.37 | 76.30 | 40.42 | 76.03 | 39.08 | 77.14 |
| 10 | 5/3/2024 12:45:30 AM | 39.71 | 76.16 | 39.74 | 74.66 | 40.27 | 76.12 | 40.12 | 74.96 | 39.37 | 76.32 | 40.46 | 75.90 | 39.08 | 77.14 |

**Figure 10: Data Report of Stability Chamber**

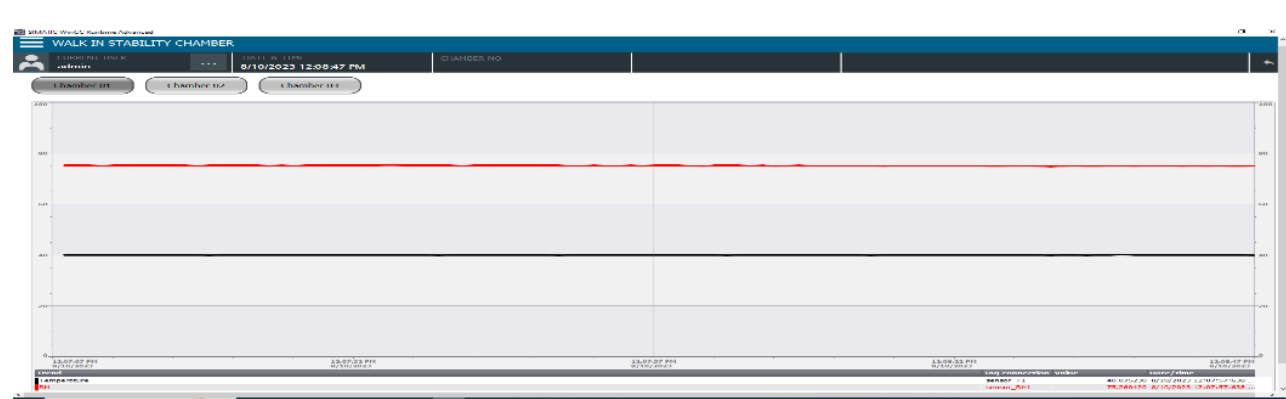

**Figure 11: Real-time Graphical Plot of Temperature and Humidity**

chilled water as needed. The capacity of the cooling coil is 2.5 TR which ensures precise temperature control. Meanwhile, the heater boasts a capacity of 4.2 kW for efficient heating of the working fluid. Controlled by a solid-state relay, the heater optimizes power usage. The stability testing chamber is located within an environment set to a controlled temperature of 24℃.

The relative humidity is controlled by a steam generator located at the relative humidity is regulated by a steam generator located at the service chamber at the rear end of the stability testing chamber. Upon activation prompted by the feedback system of the stability chamber, the steam produced by the generator is directed through a perforated pipe into the air-steam mixing zone. Processed air from the cooling coil and heater combines with the steam as per the demand of the chamber. Subsequently, the air-steam mixture is drawn by the blower and circulated into the chamber to control the environment. The heater of the steam generator is governed by a current control valve, regulating the current range between 0 A to 25 A through the analog method. With a steam output capacity of 7.2 liters per hour, the steam generator adequately maintains the desired relative humidity levels within the chamber. Each cooling coil, heater, and steam generator (Cooling Coil 1, Heater 1, and Steam Generator 1) is paired with a backup cooling coil, heater, and steam generator (Cooling Coil 2, Heater 2, and Steam Generator 2) to ensure continuous operation in the event of a breakdown of the main system. The placement of the cooling coil facilitates easy separation of condensate, directing it smoothly into the drainage system. Designed with compactness and professionalism in mind, the stability testing chamber offers a volume of approximately 10,000 liters within a modest footprint of just 53 square feet. Figure 4 depicts the practical implementation of the proposed system. The system was designed according to the schematics described previously.

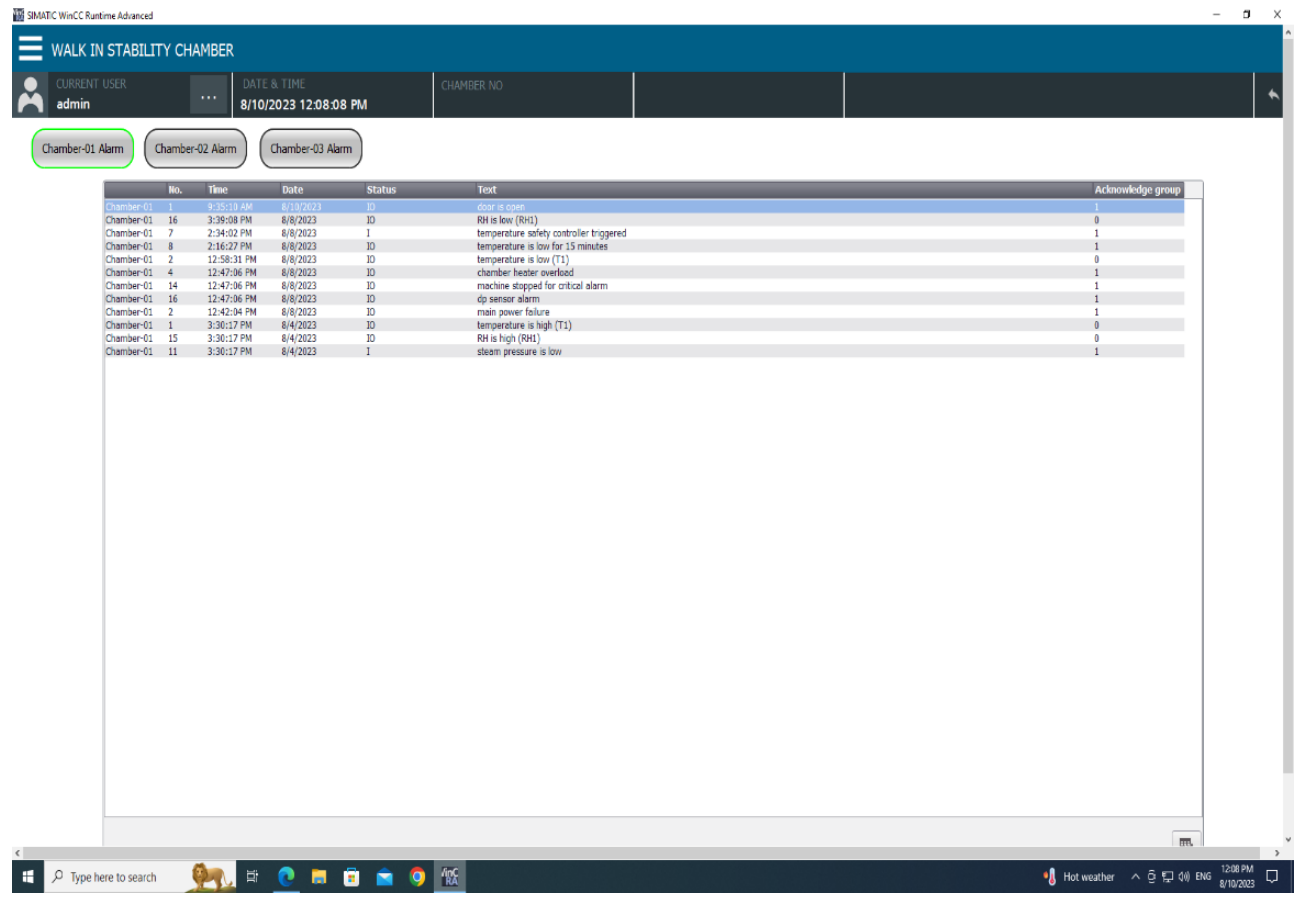

**Figure 12: Alarm Management Screen**

## 4 Electrical System Development

The utility system of the stability chamber is managed by a control system based on a Programmable Logic Controller (PLC), with the Siemens S7-1200 PLC acting as the primary control unit, as shown in Figure 5.

To control the chamber's humidity, the steam generator is controlled by a power controller, HANYOUNG NUX TPR-2N, that controls the current of the water heater of the steam generator using analog PID control. Through pre-tuning and fine-tuning phases, PID parameters were established for PID control, and sensor readings were used as the feedback of the system. The system maintains a tolerance of ±5% humidity and ±2°C temperature. The control system of actual implementation is depicted in Figure 6. The specifications of the electrical components are listed in Table 2.

The magnetic contactor (Siemens 3RT2025-1BB40) handles the switching of the 3-phase 440V output, controlled by a 24VDC coil signal from the PLC. To ensure safety and manage power integrity, ABB SH203-C25 circuit breakers are used to protect against overloads and short circuits. A centrifugal blower facilitates air circulation within the stability chamber. A Dwyer pressure transmitter is utilized to monitor pressure within the chamber. The pressure reading indicates the state of the blower, whether it is running or not, and provides an alarm if the blower fails to function. All the sensors' values can be retrieved from PLC and are monitored using a user interface.

## 5 User Interface Development

The user interface was developed using the Siemens WinCC Runtime Advanced V15.1 visualization platform, which includes capabilities like data monitoring, user authentication, report printing, and alarm history in accordance with the FDA 21 CFR Part 11 regulation, the front end for system administration is designed. Figure 7 depicts the monitoring interface of the chambers where the temperature and relative humidity can be monitored from a single screen which is convenient for the users.

The user list screen is shown in Figure 8 (some parts are blurred to protect user information). There are three user levels for user management: Administrator, Supervisor, and Operator. Administrators have the highest level of access to the system and can modify different critical parameters whereas Operators have only the access to monitor the temperature value and relative humidity (RH). Supervisors can set the parameters of the temperature and relative humidity and monitor the system parameters and sensor reading. A script is developed that is integrated with the user interface that has the capability to send the monitored data to the Microsoft SQL database. Historical data reports can be generated using Microsoft SQL Server Reporting Service (SSRS).

## 6 IoT System Development

A Python-based application has been developed to retrieve PLC data, using the CTYPES Python Wrapper for SNAP7 which can interface with Siemens S7 PLCs, including newer CPU types like the S7-1200 and S7-1500. SNAP7 is an open-source, multi-platform Ethernet communication suite developed for seamless interaction with Siemens S7-PLCs [12]. Real-time data monitoring is achieved using the MQTT protocol, ensuring minimal latency for any IoT applications. MQTT's efficiency, with a smaller header compared to HTTP, optimizes performance in bandwidth-limited environments. The real-time data is sent to the public Mosquitto MQTT broker to a specific topic using JSON format. Any client can subscribe to a specific topic for real-time data updates. Subscribers receive updated data from the broker periodically. Additionally, a Django 3.4.1 Restful API has been implemented to store data on cloud servers. Authorized users can access data via a website or any client application through the broker. The IoT architecture of the proposed system is depicted in Figure 9.

**Table 2: List of electrical components**

| Component | Specifications |
|---|---|
| PLC | Brand: Siemens, Model: S7-1200 |
| Temperature and Humidity Sensor | Brand: Rotronic, Model: HF132-WB1XX1XX |
| Tubular heater with fins | Quantity: 1000Watts X 6 pieces |
| Solid State Relay | Brand: Fotek<br>Model: SSR-40<br>Spec: Input: 3-32 VDC<br>Output: 24-380 VAC |
| Power Supply | Brand: Schneider electric<br>Model: RXM<br>Spec: Output 100-240 VAC,<br>2.8-1.4 A<br>Input 24VDC, 5A, 120W |
| Power Controller | Brand: HANYOUNG NUX<br>Model: TPR-2N<br>Spec: Source 220V, 50-60 HZ<br>Ratings: 25 A<br>Input: 4-20 mA |
| Magnetic Contactor | Brand: Siemens<br>Model: 3RT2025-1BB40<br>Spec: Coil - 24 VDC<br>Output: 3 Phase 440V |
| Circuit Breaker | Brand: ABB<br>Model: SH203-C25<br>Spec: 3Phase, 25A |
| Pressure Transmitter | Brand: Dwyer<br>Model: 616KD-B-04-AT<br>Range: 0-10 WC |
| Single Phasing Preventor | Brand: Mimilec<br>Model: VSP 3<br>Spec: 415 VAC, 50 Hz, NO |
| Blower | Brand: EBM-PAPST<br>Type: Centrifugal blower<br>Power supply: 300W, 1.31A<br>Model no: D2E146-AP47-97 |

## 7 Results

The developed system in this paper has demonstrated robust performance throughout the running phase. Key parameters like temperature and humidity have been monitored continuously and controlled with high efficiency. The data report from the SSRS shows that sensor data is within the range (Figure 10).

The system has the feature to view the real-time graphical plot of the temperature and the relative humidity (Figure 11). Any error in the system can also be monitored using the alarm management feature of the system.

The system notifies the user regarding the using email and it can also be viewed using the alarm screen (Figure 12). As previously described, the IoT architecture utilizes SNAP7, wrapper to directly retrieve and handle the temperature and relative humidity data from the PLC. Data is then transmitted to a public Mosquitto Message Queuing Telemetry Transport (MQTT) broker using JSON format [13]. To test the functionality of the IoT feature, MQTTfx, a PC-based MQTT client, was used. The MQTT client is successfully receiving the data directly from the PLC via MQTT broker. The output of the MQTT client is depicted in Figure 13.

Along with the state-of-the-art features, this stability chamber is cost-effective than traditional stability chamber in the market with the same capacity. The cost of development of each chamber is 60% cheaper than the regular stability chamber as it was developed in-house. Also, the cooling system of this chamber uses in-house chilled water whereas regular stability chambers use refrigerant-based cooling system that incurs additional electricity cost which is avoided by the developed system.

## 8 Conclusion

In conclusion, the developed stability testing chamber presented in this paper demonstrates effective control and environmental monitoring (Temperature and Relative Humidity) across four distinct chambers. The integration of advanced technologies such as PLC-based control, PID algorithms, and IoT architecture to retrieve data directly from PLC has enhanced the system's capability in monitoring the sensor's readings. The user interface, compliant with the FDA 21 CFR Part 11 guidelines, offers comprehensive data visualization and management capabilities, ensuring user-friendly operation. However, there are scopes of improvement in this work with the implementation of machine learning in this system to improve the accuracy of the sensor reading and detect the failure of the system before breakdown. Additionally, the IoT capability is tested using MQTTfx software and in the future, a complete client application will be developed which will include advanced data analytics and more diverse control capabilities.

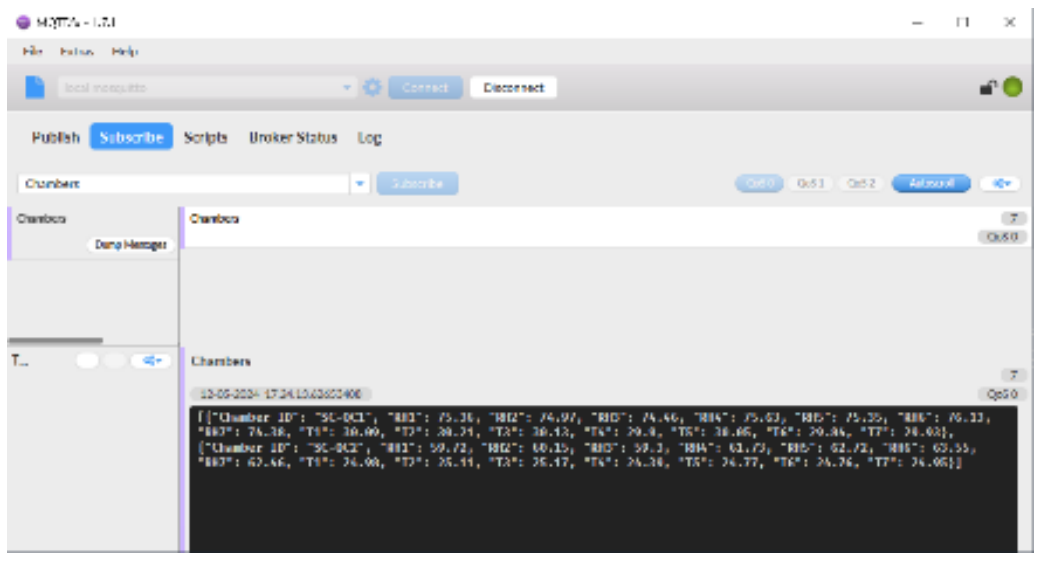

**Figure 13: Output of the MQTT Client**

## Acknowledgments

This research project is funded by Renata Limited, Bangladesh.

While preparing this article, the author(s) used "ChatGPT" to improve the grammar correction and readability of some contents of the paper. After using this tool, the author(s) reviewed and edited the content as needed.